\documentclass[superscriptaddress,secnumarabic,
amssymb,amsmath,nobibnotes,aps,prd,showkeys,showpacs,nofootinbib]{revtex4}%
\usepackage{graphicx}
\usepackage{float}
\usepackage{epsf}
\usepackage{bm}
\usepackage{amsmath}
\usepackage{amsfonts}
\usepackage{amssymb}
\usepackage{epstopdf}
\usepackage{natbib}
\usepackage{color}%
\providecommand{\U}[1]{\protect\rule{.1in}{.1in}}
\newcommand{\be}{\begin{equation}}
\newcommand{\ee}{\end{equation}}

\newcommand{\mincir}{\raise
-3.truept\hbox{\rlap{\hbox{$\sim$}}\raise4.truept\hbox{$<$}\ }}
\newcommand{\magcir}{\raise
-3.truept\hbox{\rlap{\hbox{$\sim$}}\raise4.truept\hbox{$>$}\ }}

\newtheorem{remark}{Remark}[section]
\newcommand {\R}{\mathbb{R}}

\begin{document}

\title{{ On the Hole Argument and the physical interpretation of General Relativity}}

\author{Jaume de Haro\footnote{E-mail: jaime.haro@upc.edu}}
\affiliation{Departament de Matem\`atiques, Universitat Polit\`ecnica de Catalunya, Diagonal 647, 08028 Barcelona, Spain}

\begin{abstract}

Einstein presented the Hole Argument against General Covariance, understood as invariance with respect to a change of coordinates, as a consequence of his initial failure to obtain covariant equations that, in the weak static limit, contain Newton's law. Fortunately, about two years later, Einstein returned to General Covariance and found these famous equations of gravity. However, the rejection of his Hole Argument carries a totally different vision of space-time. Its substantivalism notion, which is an essential ingredient in Newtonian theory and also in his special theory of relativity, has to be replaced, following Descartes and Leibniz's relationalism, by a set of "point-coincidences."

\end{abstract}

\vspace{0.5cm}


\maketitle

\thispagestyle{empty}

\section{Introduction}

The Einstein Hole Argument is a notable and intriguing challenge to General Covariance. This argument, presented by Einstein in 1913, supports his rejection of a general covariant (covariant with respect change of co-ordinates) theory of gravitation, because  in his initial attempts well before 1915, he was unable to construct a general covariant theory that, for static weak gravitational fields, would yield the Newtonian theory.

\

At its core, the Hole Argument challenges the conventional \textit{substantivalism doctrine}, which entails understanding space-time as an objective entity with intrinsic properties independent of its content. This raises questions about the individuality of space-time points and whether one can attribute distinct physical characteristics to them. Einstein's original argument, adopting a substantivalism point of view, explores the tension between the co-ordinate freedom inherent in general covariant theories and the uniqueness of their solutions.

\

In this context, the Hole Argument emerges as a thought experiment that probes the consequences of the mathematical structure of the theory on our conceptualization of physical reality. By scrutinizing the implications of General Covariance, the Hole Argument prompts us to reconsider our intuitions about the nature of space-time and grapple with the profound interplay between mathematical formalism and physical interpretation. Its resolution has far-reaching implications for our understanding of the ontology of space-time and continues to stimulate philosophical discussions within the realm of General Relativity \cite{Stachel, Earman-Norton, Pooley, Pooley-Read, Weatherall, Norton, Landsman,Norton1}, although some of these discussions seem to me to take on a metaphysical character, which can lead to losing the physical content of the problem. In the same sense, sometimes the abuse and misuse of complicated mathematical structures obscure the reasoning and make it difficult to understand their real physical meaning.

\

{ In fact, as we will discuss throughout the work, there exists a disconnect between Einstein's Hole Argument against General Covariance and the modern iteration used by some authors, where the covariance or lack thereof of the stress-energy tensor does not play a substantial role. On the contrary, the new version constitutes a discussion, albeit sometimes concealed and poorly understood, about the diffeomorphism invariance of the field equations. The true essence of some of these papers lies in the implicit assumption of Gauge Covariance, leading to the trivial observation that the field equations possess infinitely many solutions. Consequently, the Hole Argument emerges as a simple corollary devoid of any essential significance.

Once the trivial link between the assumed Gauge Covariance and the lack of uniqueness of solutions is established, albeit through unnecessarily convoluted mathematical tools, a philosophical debate ensues. This debate, contrary to physical intuition, revolves around the mathematical viewpoint—especially the geometric aspect—of the field equations and spacetime structure, sometimes resulting in erroneous physical interpretations of General Relativity.

\

}

\

From a physical approach to the problem, the main goal of the present {essay} is, following Rovelli's reasoning in \cite{Rovelli}, to explain the most agreed-upon point of view in the physics community regarding the resolution of the Hole Argument. It can be summarized as follows: To preserve General Covariance, one has to reject substantivalism of space-time, meaning the rejection of absolute space-time, and accept the \textit{relationalistic doctrine}. In this perspective, events in space-time can be defined "à la Einstein" as coincidence points of world-lines. Adopting this viewpoint, the Hole Argument only shows that a deterministic covariant theory demands the abandonment of absolute space-time, where events have an intrinsic physical meaning independently of what surrounds them.

\

However, one has to be cautious with the fact that sometimes physicists argue that in General Relativity, "the space-time is dynamic" in the sense that it interacts with the matter contained within it, providing the space-time with a metric. This could lead to the temptation to attribute some individuality to the events of space-time, as in Newtonian absolute space. However, in General Relativity, one cannot dissociate space-time from matter fields and other fields involved in the theory; that is, points are related to fields. Following this line of thinking, one can argue that positions are defined with respect to fields via the coincidence of world-lines or geodesics whose nature is intrinsically related to the metric, and thus with the fields that provide it.

\

The work is organized as follows: In Section II, we provide a review of pseudo-Riemann geometry, demonstrating that diffeomorphisms preserve the intrinsic curvatures of a manifold. The presentation of Einstein's equation in this intrinsic form is given in Section III, where we also discuss the Hole Argument as a challenge to tensor field equations. In that section, we also introduce the concept of Gauge Covariance (invariance by diffeomorphisms), eliminating the need for a Hole to present arguments for or against substantivalism. Section IV is dedicated to the presentation of the original Einstein Hole Argument against general covariance, i.e., invariance with respect to the change of coordinates or, in modern language, invariance with respect to passive diffeomorphisms. This section includes Hilbert's elaboration of the Hole Argument and its most popular interpretation in the physics community. In Section V, I attempt to shed light on works with a profound philosophical essence and a highly complex mathematical foundation. Finally, in the last Section, I present my conclusions.

\section{Brief review of pseudo-Riemann manifolds}

The objective of this section is to provide a foundational understanding of differential geometry, establishing the necessary groundwork to navigate tensor equations within a manifold.

\

Let me start with 
an $n+m$-dimensional pseudo-Riemann manifold  represented by a pair $(\mathcal{M}, \mathfrak{g})$, where it is possible to select a \textit{frame field}  ${E_1, \ldots, E_n, E_{n+1}, \ldots, E_{n+m}}$ in 
$\mathfrak{X}(\mathcal{M})$ (the set of vector fields in $\mathcal{M}$ or in a more technical language the  sections of the 
the tangent bundle $T\mathcal{M}$) that satisfies:
\begin{eqnarray}
\mathfrak{g}(E_1,E_1)= \ldots =\mathfrak{g}(E_n,E_n)=1, \quad
\mathfrak{g}(E_{n+1},E_{n+1})= \ldots =\mathfrak{g}(E_{n+m},E_{n+m})=-1,
\end{eqnarray}
and $\mathfrak{g}(E_i,E_i)=0$ when $i\neq j$.

\

Consider an \textit{active} diffeomorphism (a diffeomorphism different from the identity), denoted as $\phi$, from $\mathcal{M}$ to itself. This induces an isometry defined by
\begin{eqnarray}
\Phi: (\mathcal{M}, \widetilde{\mathfrak{g}}) \longrightarrow (\mathcal{M}, \mathfrak{g}),
\end{eqnarray}
where the metric $\widetilde{\mathfrak{g}}$ is the pull-back of $\mathfrak{g}$, expressed as $\widetilde{\mathfrak{g}}=\phi^*\mathfrak{g}$. Effectively, for two vector fields $X$ and $Y$ in $\mathfrak{X}(\mathcal{M})$, we have
\begin{eqnarray}
\widetilde{\mathfrak{g}}(X,Y)=\mathfrak{g}(\phi_*X,\phi_*Y),
\end{eqnarray}
where $\phi_*Z$ represents the push-forward of a vector $Z$. This push-forward is defined in the usual manner, as $\phi_*Z(f)=Z(\phi^*f)=Z(f\circ \phi)$.

\

Since isometries preserve angles and distances, one can argue mathematically that the pseudo-Riemann manifolds $(\mathcal{M}, \widetilde{\mathfrak{g}}=\phi^*\mathfrak{g})$ and $(\mathcal{M}, \mathfrak{g})$ are identical. More precisely, any \textit{measure} (for example, the scalar product of two vector fields or the calculation of the scalar curvature), denoted as $\widetilde{E}$, in $(\mathcal{M}, \widetilde{\mathfrak{g}})$ at the point $p\in \mathcal{M}$, when mapped by $\Phi$ to the \textit{measure} $E$ in $(\mathcal{M}, \mathfrak{g})$ at the point $\phi(p)\in \mathcal{M}$, predicts the same result as $E$.

\

To express this fact mathematically, 
following \cite{Iftime-Stachel},
let us to  introduce the group of active diffeomorphisms and the class or orbit of a tensor as follows:
\begin{eqnarray}
    \mathcal{M}_{diff}=\{\phi: \mathcal{M}\longrightarrow
    \mathcal{M}, \quad \phi \mbox{ diffeomorphism}\},
\end{eqnarray}
and given a $2$-covariant tensor $\mathfrak{H}: \mathfrak{X}(\mathcal{M})\times \mathfrak{X}(\mathcal{M})\longrightarrow \mathcal{C}^{\infty}(\mathcal{M})$,  we define its class or orbit by
\begin{eqnarray}
    [\mathfrak{H}]=\{\phi^*\mathfrak{H}, \quad \forall \phi \in \mathcal{M}_{diff}\}.\end{eqnarray}

These orbits are the elements of the  topological quotient space 
$\mathfrak{Q}_2(\mathcal{M})=\mathfrak{TEN}_2/\mathcal{M}_{diff}$, where $\mathfrak{TEN}_2$ is the vector space of all symmetric $2$-covariant tensors. 
So, given a metric $\mathfrak{g}^0$ in $\mathcal{M}$
we can say that its orbit $[\mathfrak{g}^0]
\in\mathfrak{Q}_2(\mathcal{M})$ contains all the 
metrics mathematically equivalent to $\mathfrak{g}^0$ in $\mathcal{M}$.

\

\

Next, we consider the Levi-Civita connection of the  metric $\mathfrak{g}$, namely $\nabla^{\mathfrak{g}}$, 
which is obtained imposing the  symmetry condition or torsion free: 
\begin{eqnarray}
    \nabla^{\mathfrak{g}}_XY-\nabla^{\mathfrak{g}}_YX=[X,Y],
\end{eqnarray}
where $[X,Y]=XY-YX$ is the Lie bracket, and the compatibility  with $\mathfrak{g}$, i.e.,
\begin{eqnarray}
    \nabla^{\mathfrak{g}}\mathfrak{g}=0,
\end{eqnarray}
where the covariant derivative of a $2$-covariant tensor is defined as 
\begin{eqnarray}\nabla^{\mathfrak{g}}_Z\mathfrak{H}(X,Y)=
    Z\mathfrak{H}(X,Y)-\mathfrak{H}(\nabla^{\mathfrak{g}}_ZX,Y)-\mathfrak{H}(X,\nabla^{\mathfrak{g}}_ZY).
\end{eqnarray}

\

The fundamental property of the Levi-Civita connection, which comes from the uniqueness of this connection,  is that (see the Proposition 5.6 of \cite{Lee})
\begin{eqnarray}\label{Levi}
   \phi^*\nabla^{\mathfrak{g}}= \nabla^{\widetilde{\mathfrak{g}}}
\Longrightarrow\phi_*(\nabla^{\widetilde{\mathfrak{g}}}_XY)=\nabla^{\mathfrak{g}}    _{\phi_*X}\phi_*Y.
\end{eqnarray}

\

Next we introduce the Riemann tensor. First of all we introduce the Riemann curvature endomorphism $\mathfrak{R}_{\mathfrak{g}}:
\mathfrak{X}(\mathcal{M})\times\mathfrak{X}(\mathcal{M})\times\mathfrak{X}(\mathcal{M})\rightarrow\mathfrak{X}(\mathcal{M})$ defined by
\begin{eqnarray}
  \mathfrak{R}_{\mathfrak{g}}(X,Y)Z= \nabla^{\mathfrak{g}}_X
  \nabla^{\mathfrak{g}}_Y Z-  \nabla^{\mathfrak{g}}  _Y\nabla^{\mathfrak{g}}_X Z-\nabla^{\mathfrak{g}}_{[X,Y]}Z.\end{eqnarray}

With this endomophism we can define the Riemann curvature tensor for the metric $\mathfrak{g}$, namely $\mathfrak{Rm}_{\mathfrak{g}}$,  as the $4$-covariant tensor satisfying
\begin{eqnarray}
    \mathfrak{Rm}_{\mathfrak{g}}    (X,Y,Z,W)=\mathfrak{g}(\mathfrak{R}_{\mathfrak{g}}    (X,Y)Z, W).
\end{eqnarray}

Since $\Phi$ is an isometry, because $\widetilde{\mathfrak{g}}=\phi^* \mathfrak{g}$, one has the fundamental property (see the Lemma 7.2 of \cite{Lee}) \begin{eqnarray}\label{fundamental}
    \phi^* \mathfrak{Rm}_{\mathfrak{g}}    ={\mathfrak{Rm}}_{\widetilde{\mathfrak{g}}},
\end{eqnarray}
which comes from (\ref{Levi}) and tells us that the Riemann tensor is invariant for diffeomorphisms.

\

Once we have defined the curvature we can define the $2$-covariant Ricci tensor as the trace of the Riemann curvature endomorphism, as follows:
\begin{eqnarray}
    \mathfrak{Rc}_{\mathfrak{g}}    (X,Y)=Tr( Z\rightarrow \mathfrak{R}_{\mathfrak{g}}    (Z,X)Y).
\end{eqnarray}

To calculate this trace we define the 
{\it flat} operator 
$\flat: \mathfrak{X}(\mathcal{M})\rightarrow \mathfrak{X}^*(\mathcal{M})$, defined as follows:
\begin{eqnarray}
    \flat: X\rightarrow X^{\flat}, \quad \mbox{where} \quad X^{\flat}(Y)=\mathfrak{g}(X,Y),
\end{eqnarray}
and its inverse $\sharp=\flat^{-1}$ named {\it sharp} operator.

Let now $E^i$ be the dual of the vector $E_i$, that is $E^i(E_j)=\delta_{ij}$, being $\delta_{ij}$ the Kronecker's delta. We want to calculate ${E^i}^{\sharp}$. Since the sharp is the inverse of the flat,  we will have:
\begin{eqnarray}
    \mathfrak{g}({E^i}^{\sharp}, E_j)
    ={{E^i}^{\sharp}}^{\flat}(E_j)=
    E^i(E_j)=\delta_j^i,\end{eqnarray}
meaning that 
\begin{eqnarray}
    {E^i}^{\sharp}= \mathfrak{g}^{-1}(E^i, E^k)E_k,
    \end{eqnarray}
where we have used the Einstein's summation convention and we have introduced the $2$-contravariant tensor $\mathfrak{g}^{-1}:\mathfrak{X}^*(\mathcal{M})\times \mathfrak{X}^*(\mathcal{M})\longrightarrow \mathcal{C}^{\infty}(\mathcal{M})$ as the inverse of $\mathfrak{g}$, i.e., $\mathfrak{g}^{-1}(E^i, E^k)
\mathfrak{g}(E_k, E_j)=\delta_j^i$.
Then, the Ricci tensor can be calculated as follows:
\begin{eqnarray}
    \mathfrak{Rc}_{\mathfrak{g}}    (X,Y)= \mathfrak{Rm}_{\mathfrak{g}}    (E_i,X,Y,{E^i}^{\sharp}),
\end{eqnarray}
and, from the property (\ref{fundamental}) we have
\begin{eqnarray}
\phi^* \mathfrak{Rc}_{\mathfrak{g}}=
{\mathfrak{Rc}}_{\widetilde{\mathfrak{g}}}.
\end{eqnarray}

\

To define the scalar curvature (referred to as the Ricci scalar in physics), denoted as 
$R$, as the trace of the Ricci tensor, we first need to introduce the concept of the trace of a 
$2$-covariant tensor $\mathfrak{H}: \mathfrak{X}(\mathcal{M})\times \mathfrak{X}(\mathcal{M})\rightarrow \mathcal{C}^{\infty}(\mathcal{M})$, which we can be defined  as follows
\begin{eqnarray}
    Tr_{\mathfrak{g}}\mathfrak{H}=\mathfrak{H}(E_i, {E^i}^{\sharp}),
    \end{eqnarray}
independently of the  chosen basis.  

Then,  applying this definition to the Ricci tensor we will have 
\begin{eqnarray}
    R_{\mathfrak{g}}= \mathfrak{Rc}_{\mathfrak{g}}    (E_i,{E^i}^{\sharp}).
\end{eqnarray}

\

In summary, 
the lesson we have to learn  here is that the pull-back preserves the intrinsic curvatures of the manifold, in the sense that
\begin{eqnarray}\label{invariance}
    \phi^*: (\mathfrak{Rm}_{\mathfrak{g}},\mathfrak{Rc}_{\mathfrak{g}},  R_{\mathfrak{g}})\rightarrow ({\mathfrak{Rm}}_{\widetilde{\mathfrak{g}}}    ,{\mathfrak{Rc}}_{\widetilde{\mathfrak{g}}}, {R}_{\widetilde{\mathfrak{g}}}).
\end{eqnarray}

\section{Einstein equation}

With all this machinery, one can introduce the $2$-covariant Einstein's equation in $\mathfrak{TEN}_2$,  in its intrinsic form:
\begin{eqnarray}\label{Einstein}
   \mathfrak{G}_{\mathfrak{g}}\equiv \mathfrak{Rc}_{\mathfrak{g}}-\frac{1}{2}R_{\mathfrak{g}}\mathfrak{g}=\kappa\mathfrak{T}_{\mathfrak{g}},
\end{eqnarray}
where $\mathfrak{G}_{\mathfrak{g}}$ is the so-called Einstein tensor
and
$\mathfrak{T}_{\mathfrak{g}}$ is the $2$-covariant stress-energy tensor,  for example,  the one corresponding to a perfect fluid
\begin{eqnarray}
    \mathfrak{T}_{\mathfrak{g}}=\left(\rho+\frac{p}{c^2}\right)\mathfrak{u}\otimes\mathfrak{u}+p\mathfrak{g},
\end{eqnarray}
where $\rho$ is the energy density of the fluid, $p$ is its pressure and $\mathfrak{u}$ is the vector field of $4$-velocities.

In this context, $\kappa=\frac{8\pi G}{c^4}$, where $G$ represents Newton's constant, and $c$ signifies the velocity of light. It is crucial to grasp that the metric $\mathfrak{g}$ is the unknown in this equation. According to Einstein's "Mach principle," the mass of bodies dictates geodesic motion. This concept serves as a criterion for selecting solutions to Einstein's equation \cite{Raine}. In this sense, the stress-energy tensor $\mathfrak{T}_{\mathfrak{g}}$ determines the metric $\mathfrak{g}^0$, the sole "physically meaningful" solution to the Einstein equation (\ref{Einstein}), thereby establishing the pseudo-Riemann manifold $(\mathcal{M},\mathfrak{g}^0)$.

\

\begin{remark}
{ In a $4$-dimensional Lorentz manifold 
 ($\mathfrak{g}(E_4,E_4)=-1$ and    $\mathfrak{g}(E_j,E_j)=1$ for $j=1,2,3$), one has $Tr_{\mathfrak{g}}(\mathfrak{g})= g(E_i, {E^i}^{\sharp})=4$, and thus,  taking the trace in both sides of Einstein's equation one has $R_{\mathfrak{g}} =-\kappa T_{\mathfrak{g}}$  where $T_{\mathfrak{g}} =Tr_{\mathfrak{g}} (\mathfrak{T}_{\mathfrak{g}})$. Therefore, the Einstein equation can also be written as 
 \begin{eqnarray}
     \mathfrak{Rc}_{\mathfrak{g}}=\kappa\mathfrak{T}-\frac{\kappa}{2}T_{\mathfrak{g}}     \mathfrak{g}.
 \end{eqnarray}}
\end{remark}

\

Next we consider an active diffeomorphism $\phi: \mathcal{M}\rightarrow\mathcal{M}$ which pulls-back the stress-tensor $\mathfrak{T}_{\mathfrak{g}}$ to $\phi^* \mathfrak{T}_{\mathfrak{g}}$. We have 
$\phi^*\mathfrak{G}_{\mathfrak{g}}= 
{\mathfrak{G}}_{\phi^*\mathfrak{g}}$, but in general $\phi^*\mathfrak{T}_{\mathfrak{g}}\not= {\mathfrak{T}}_{\phi^*\mathfrak{g}}$, and the question is to compare the Einstein equations
\begin{eqnarray}
 \mathfrak{G}_{\mathfrak{g}}  
    = \kappa\mathfrak{T}_{\mathfrak{g}}
    \qquad \mbox{and}  \qquad
 {\mathfrak{G}}_{{\phi^*\mathfrak{g}}}  
=\kappa\phi^*\mathfrak{T}_{\mathfrak{g}}.
\end{eqnarray}

It is clear that if $\mathfrak{g}^0$ is the solution (we will assume uniqueness) of 
$\mathfrak{G}_{\mathfrak{g}} 
    = \kappa\mathfrak{T}_{\mathfrak{g}}$, then 
    $\mathfrak{\widetilde{g}}^0=  \phi^*\mathfrak{g}^0$
is only a solution of 
$\mathfrak{G}_{\mathfrak{g}}  
    = \kappa\phi^*\mathfrak{T}_{\mathfrak{g}}$, 
    provided that 
    $\phi^*\mathfrak{T}_{\mathfrak{g}}= {\mathfrak{T}}_{\phi^*\mathfrak{g}}$.

Therefore, it is important to stress that when 
$\phi^*\mathfrak{T}_{\mathfrak{g}}\not={\mathfrak{T}}_{\phi^*\mathfrak{g}}$, the diffeomorphism $\phi$ does not induce another  solution of the Einstein equation.

\

\

However, although in general  $\phi^*\mathfrak{T}_{\mathfrak{g}}\not={\mathfrak{T}}_{\phi^*\mathfrak{g}}$, 
there is a class of active diffeomorphisms, which we will name {\it hole-diffeomorphisms}, that pulled-back  the stress-energy tensor to itself, i.e.,  
$\phi^*\mathfrak{T}_{\mathfrak{g}}= \mathfrak{T}_{\phi^*\mathfrak{g}}$

\

To define this kind of diffeomorphism we consider an open subset $\mathcal{H}$ (the hole) of $\mathcal{M}$ where the stress-energy tensor vanishes. Then, we consider diffeomorphisms 
$\phi_h$ from $\mathcal{M}$ to itself, satisfying:
\begin{eqnarray}
    \phi_h=Id. \quad \mbox{in} \quad \mathcal{M}\setminus \mathcal{H};\qquad  \phi_h\not=Id. \quad \mbox{in} \quad \mathcal{H},
\end{eqnarray}
where $Id.$ denotes the identity. From this definition, we can see that 
$\phi_h^*\mathfrak{T}_{\mathfrak{g}}= {\mathfrak{T}}_{\phi_h^*\mathfrak{g}}$, and thus 
from a solution $\mathfrak{g}^0$  of the Einstein equation
$\mathfrak{G}_{\mathfrak{g}}
   = \kappa\mathfrak{T}_{\mathfrak{g}}$ we can construct another different one $\mathfrak{\widetilde{g}}^0=\phi_h^*\mathfrak{g}^0$.

\

Note that outside the hole, both solutions coincide, but inside it, they differ. This is the foundation of the "hole argument" against tensor field equations. One way to express it is as follows:

\

{\it Assuming tensor field equations in the form of a geometric tensor constructed by intrinsic curvatures of the manifold being proportional to the stress-energy tensor, the lack of uniqueness in solutions introduces indeterminism.}

\

Now, there are two distinct positions we can take to address this assertion:

\begin{enumerate}
\item Accepting that the field equation doesn't necessarily have to be a tensor equation.
\item
Abandoning substantivalism and embracing a relationalistic approach to define events on the manifold. This allows us to interpret different solutions provided by a hole-diffeomorphism as representing the same physical reality.

\end{enumerate}
In the following sections, we will explore why the latter option is the correct one for formulating a theory of Gravity, including General Covariance."

\subsection{Gauge Covariance}\label{gauge-invariance}

We can extend our exploration by imposing Gauge Covariance, signifying invariance under diffeomorphism (for a more schematic definition, refer to \cite{Landsman}). The formulation can be expressed as follows:

Let's assume the presence of fields $\mathfrak{A}_1$, ..., $\mathfrak{A}_k$ within a physical system, adhering to tensor equations of the following form:

\begin{eqnarray}
\mathfrak{B}_1(\mathfrak{g},\mathfrak{A}_1,..,\mathfrak{A}_k)=0,..., \mathfrak{B}_m
(\mathfrak{g},\mathfrak{A}_1,..,\mathfrak{A}_k)=0.
\end{eqnarray}

The complete dynamical system includes the Einstein equation and the equations satisfied by the fields, i.e.,
\begin{eqnarray}\label{dinamical}
  \mathfrak{G}_{\mathfrak{g}}=\kappa\mathfrak{T}_{\mathfrak{g}}(\mathfrak{A}_1,...,\mathfrak{A}_k),\quad\mathfrak{B}_1(\mathfrak{g},\mathfrak{A}_1,..,\mathfrak{A}_k)=0,..., \mathfrak{B}_m
(\mathfrak{g},\mathfrak{A}_1,..,\mathfrak{A}_k)=0.
\end{eqnarray}
Then, given a diffeomorphism $\phi$, a Gauge Covariant theory transform  the equations (\ref{dinamical}) via the pull-back $\phi^*$ as follows:
\begin{eqnarray}\label{dinamical1}
  \mathfrak{G}_{\mathfrak{\phi^*g}}=
  \kappa\mathfrak{T}_{\phi^*\mathfrak{g}}(\phi^*\mathfrak{A}_1,...,\phi^*\mathfrak{A}_k),\quad\mathfrak{B}_1(\phi^*\mathfrak{g},\phi^*\mathfrak{A}_1,..,\phi^*\mathfrak{A}_k)=0,..., \mathfrak{B}_m
(\phi^*\mathfrak{g},\phi^*\mathfrak{A}_1,..,\phi^*\mathfrak{A}_k)=0.
\end{eqnarray}

Therefore, let $\mathfrak{R}(\mathcal{M})$ the space of solutions of (\ref{dinamical}). Given a solution $\{\mathfrak{g}^0,\mathfrak{A}_1^0,..,\mathfrak{A}_k^0\}$ of (\ref{dinamical}) , for each diffeomorphism we obtain another solution
$\{\phi^*\mathfrak{g}^0,\phi^*\mathfrak{A}_1^0,..,\phi^*\mathfrak{A}_k^0\}$. That is, we obtain a class of solutions  $[\mathfrak{g}^0,\mathfrak{A}_1^0,..,\mathfrak{A}_k^0]$
belonging to $\mathfrak{R}(\mathcal{M})/\mathcal{M}_{diff}$.

\

In a Gauge Covariant theory, one that remains invariant under diffeomorphism, the resolution to eliminate indeterminacy requires acknowledging that all elements within a class of solutions represent the same physical state (refer to Chapter 2 of \cite{Rovelli}).

This acknowledgment entails abandoning substantivalism and embracing the Earman-Norton 'Leibniz Equivalence,' as articulated in \cite{Davis} where it is stated, 'Diffeomorphic models represent the same physical situation.' This shift involves adopting a relationalistic approach to define events on the manifold.

A final point worth noting:

\begin{remark}\label{Remark}
Embracing Gauge Covariance obviates the need for a Hole to argue against covariance. Given a solution to the equations, all diffeomorphisms applied to this solution lead to solutions belonging to the same class, resulting in indeterminism without the need for the Hole. This indeterminacy, naturally, dissipates when rejecting substantivalism in favor of Einstein's relational viewpoint.
\end{remark}

\

\section{Einstein's hole argument}\label{Einstein_hole}

The history of the Hole Argument comes from its initial failure to obtain General Covariant field equations that in the static weak limit contain the Newtonian equations. I direct to the reader to papers about the foundation of General Relativity \cite{Norton, Norton2,Norton3,Straumann,Weinstein, Sauer}.
In this section we will present the original Hole Argument against General Covariance (covariance with respect change of co-ordinates or in more modern lenguage with respect to {\it passive} diffeomorphisms) applied to the Einstein equation, but it also holds for other tensor equations relating 
some intrinsic $2$-curvature tensors with the stress-energy tensor.

\

Let us consider a $4$-dimensional Lorentz manifold $(\mathcal{M}, \mathfrak{g})$, and to simplify the reasoning, we assume that there is only a chart $(\mathcal{M}, \psi)$ mapping $\mathcal{M}$ to $\R^4$ "in $\{x\}$ coordinates", that is, given an event $p\in \mathcal{{M}}$, $\psi$   assigns the point  $p$ to $x\in \R^4$ with coordinates $(x^1,x^2,x^3, x^4)$ in a basis $\{ e_1, e_2, e_3, e_4\}$ defined as follow:
\begin{eqnarray}
    e_j f=\frac{\partial f}{\partial x^j},  \quad \forall  f:\R^4\rightarrow \R,
\end{eqnarray}
which can be pushed-forward to $\mathcal{M}$ by $\psi^{-1}$, obtaining a basis in $\mathfrak{X}(\mathcal{M})$
\begin{eqnarray}
    E_j=\psi^{-1}_* e_j.
\end{eqnarray}

The Einstein equation $\mathfrak{G}_{\mathfrak{g}}=\kappa\mathfrak{T}_{\mathfrak{g}}$, where I assume a simplified form without intricate fields for the stress-energy tensor (recalling that originally, Einstein dealt with static matter to derive, in the weak approximation, the Newtonian law), undergoes a pull-back operation by ${\psi^{-1}}^*$, resulting in:

\begin{eqnarray}
{\psi^{-1}}^*\mathfrak{G}_{\mathfrak{g}}(e_i,e_j)(x)=\kappa {\psi^{-1}}^*\mathfrak{T}_{\mathfrak{g}}(e_i, e_j)(x).
\end{eqnarray}

Then, 
  giving a $2$-covariant tensor $\mathfrak{H}$ and denoting 
${\psi^{-1}}^*\mathfrak{H}(e_i,e_j)(x)=
\mathfrak{H}(E_i,E_j)(p)$ by $H_{ij}({\psi^{-1}}^*\mathfrak{g},  x)$, the Einstein equations in $\{ x\}$-coordinates, becomes 
\begin{eqnarray}
    R_{ij}({g}, x)-\frac{1}{2}R({g}, x)g_{ij}(x)=T_{ij}(g, x)\Longleftrightarrow 
    G_{ij}({g}, x)=T_{ij}(g,x)    ,
\end{eqnarray}
where we have used the notation
${\psi^{-1}}^*\mathfrak{g}=g$ and 
\begin{eqnarray}
 R_{ij}({g}, x)=
 {\psi^{-1}}^*\mathfrak{Rc}(e_i,e_j)(x)=
\mathfrak{Rc}(E_i,E_j)(p),\nonumber\\
R({g}, x)=R(p), \quad 
g_{ij}( x)=
\mathfrak{g}(E_i,E_j)(p), \quad 
 T_{ij}(g, x)=
\mathfrak{T}_{\mathfrak{g}}(E_i,E_j)(p).\end{eqnarray}

The expression for the entries of the Ricci tensor in ${x}$-coordinates in terms of the Christoffel symbols 
\begin{eqnarray}\label{Christoffel}
    \Gamma_{ij}^k(x)=\frac{1}{2}g^{kl}(x)\left( \frac{\partial g_{li}}{\partial x^j}(x)+
    \frac{\partial g_{lj}}{\partial x^i}(x) 
    -\frac{\partial g_{ij}}{\partial x^l}(x)    \right),
\end{eqnarray}
are
\begin{eqnarray}
 R_{ij}({g}, x)=
 \frac{\partial \Gamma_{ij}^k}{\partial x^k}(x) 
- \frac{\partial \Gamma_{ik}^k}{\partial x^j}(x) 
+\Gamma_{ij}^k(x)\Gamma_{kl}^l(x)-
\Gamma_{il}^k(x)\Gamma_{jk}^l(x). \end{eqnarray}

\

Next, let's consider another coordinate system, denoted as $\{x'\}$, where now a homeomorphism $\psi'$ maps an event $p$ to $x'$ with coordinates $(x'^1, x'^2, x'^3, x'^4)$ in the basis $\{e_1', e_2', e_3', e_4'\}$

\begin{eqnarray}
    e'_j f=\frac{\partial f}{\partial x'^j}. 
\end{eqnarray}

This basis 
 can be pushed-forward  by $\psi'^{-1}$, obtaining in $\mathfrak{X}(\mathcal{M})$, the basis 
\begin{eqnarray}
    E'_j=\psi'^{-1}_* e'_j.
\end{eqnarray}

In these coordinates, the entries of the Ricci tensor are denoted as $R_{ij}(g', x')$, where $g'={\psi'^{-1}}^*\mathfrak{g}$, and the Christoffel symbols ${\Gamma'}_{ij}^k$ are obtained from (\ref{Christoffel}) by replacing the $\{x\}$-coordinates with the $\{x'\}$ ones and $g_{ij}$ with $g'_{ij}$. However, the entries of the stress-energy tensor take the form $T'_{ij}(g',x')$, which, in general, differs from $T_{ij}(g',x')$.

\

Now we define the passive diffeomorphism (a change of coordinates)
\begin{eqnarray}
    F:\R^4\longrightarrow\R^4, \quad F(x)=(F^1(x), F^2(x), F^3(x), F^4(x))=(\psi'\circ \psi^{-1})(x),
\end{eqnarray}
which sends $x=\psi(p)$ to $x'=\psi'(p)$.

Clearly, $F^*$ pulls back the solution of the Einstein equations in $\{x\}$-coordinates to the solution in 
$\{x'\}$-coordinates. In other words, $F^*$ transforms the solution ${g'}^0$ of $G_{ij}(g', x')=\kappa T_{ij}'(g', x')$ to the solution $g^0\equiv F^* {g'}^0$ of $G_{ij}(g, x)=\kappa T_{ij}(g,x)$.

\

This embodies Einstein's perspective on 'general covariance': the field equations in ${x}$-coordinates, $G_{ij}(g, x)=\kappa T_{ij}(g, x)$, must be transformed into the equations $G_{ij}(g', x')=\kappa T_{ij}'(g', x')$ in the ${x'}$-coordinates, where

\begin{eqnarray}
    T=F^*T'\Longleftrightarrow 
    T_{ij}(g, x)=\frac{\partial F^{k}}{\partial x^i}(x)
    \frac{\partial F^{l}}{\partial x^j}(x)T'_{kl}(g', F(x)).
    \end{eqnarray}

\

To raise the Einstein's Hole Argument, 
we can also define a {\it passive hole-diffeomorphism} as follows: consider a subset $\mathcal{H}$ of $\mathcal{M}$ where the stress-energy tensor vanishes. Consider two different co-ordinate systems, namely $\psi_h$ and $\psi'_h$, satisfying 
\begin{eqnarray}
    \psi_h=\psi'_h \quad \mbox{in} \quad
    \mathcal{M}\setminus \mathcal{H} \quad 
  \mbox{and} \quad  \psi_h\not=\psi'_h \quad \mbox{in} \quad
     \mathcal{H},    \end{eqnarray}
 define  $F_h=\psi'_h\circ \psi_h^{-1}$, and let
$H=\psi(\mathcal{H})=\psi'(\mathcal{H})$ the hole in $\R^4$. Then,
\begin{eqnarray}
    F_h=Id. \quad \mbox{in} \quad \R^4\setminus H, \qquad \mbox{and} \qquad F_h\not=Id. \quad \mbox{in} \quad  H. \end{eqnarray}

\

\

The crucial observation lies in examining the stress-energy tensor in the coordinate systems $\psi_h$ and $\psi'_h$. This leads to the relationship:

\begin{eqnarray}
{\psi_h^{-1}}^*\mathfrak{T}_{\mathfrak{g}}={{\psi'}_h^{-1}}^*\mathfrak{T}_{\mathfrak{g}}\Longrightarrow
T_{ij}'(g', x')=T_{ij}(g, x)=T_{ij}(g',x'),
\end{eqnarray}

which implies that the Einstein equations in different coordinate systems are $G_{ij}(g, x)=\kappa T_{ij}(g,x)$ and $G_{ij}(g',x')=\kappa T_{ij}(g',x')$.

Now, let ${g'}^0(x')$ be a solution in $\{x'\}$-coordinates, and $g^0(x)=F^*{g'}^0(x')$ the corresponding solution in $\{ x\}$-coordinates. We can find another solution in $\{ x\}$-coordinates simply by replacing $x'$ in ${g'}^0$ with $x$, essentially relabeling the variables. In $\{x\}$-coordinates, we then have two distinct solutions, $g^0(x)$ and ${g'}^0(x)$.

Effectively, the metric tensor satisfies the following coordinate transformation
\begin{eqnarray}
    g^0_{ij}(x)=\frac{\partial F_h^k}{\partial x^i}(x)
\frac{\partial F_h^l}{\partial x^j}(x){g'}^0_{kl}(x').
\end{eqnarray}

Thus, outside the hole $H$, we will have $g^0_{ij}(x)={g'}^0_{ij}(x')={g'}^0_{ij}(F_h(x))={g'}^0_{ij}(x)$.
However, inside the hole, $g^0_{ij}(x)\neq {g'}^0_{ij}(x')$, and by relabeling the right-hand side of this last equation, we find that inside the hole $g^0_{ij}(x)\neq{g'}^0_{ij}(x)$.

\

In conclusion, within a given system of coordinates, one can find different solutions—all induced by a passive hole-diffeomorphism—of the Einstein equations. When transported to $\mathcal{M}$ by any $\psi^*_h$, these solutions result in different metrics. For instance, a solution ${g'}^0$ and its induction by $F_h^*$, denoted as ${g}^0=F^*_h {g'}^0$, are transported in ${\mathcal{M}}$ to $\mathfrak{g'}^0= \psi_h^* {g'}^0$ and $\mathfrak{g}^0= \psi_h^* g^0$. Once again, this forms the basis of Einstein's hole argument.

\

Next,  we will see that a passive diffeomorphism $F=\psi'\circ \psi^{-1}$ induces active diffeomorphisms. Effectively, define
\begin{eqnarray}
\phi: \mathcal{M}\longrightarrow
\mathcal{M}; \quad \phi=\psi^{-1}\circ F\circ \psi=\psi^{-1}\circ \psi',
\end{eqnarray}
where clearly  $\phi\not=Id$, is satisfied.

\

Applied to a passive hole-diffeomorphism 
$F_h= \psi_h'\circ \psi_h^{-1}$,  this leads to the active hole-diffeomorphism 
$\phi_h=\psi_h^{-1}\circ \psi_h'$. Note that $\phi_h$ pulls-back $\mathfrak{g}'=\psi_h^* g'$ to $\psi_h^*g=\mathfrak{g}$, because $\phi_h^*=\psi_n^*\circ F_h^*\circ {\psi^{-1}_h}^*$, and thus,
\begin{eqnarray}
    \phi_h^*\mathfrak{g}'=(\psi_h^*\circ F_h^*\circ {\psi^{-1}_h}^*)\psi_h^* g'=
    \psi_h^*\circ F_h^* g'
    =\psi_h^* g=\mathfrak{g}.\end{eqnarray}

Therefore, $\phi_h$ induces the isometry
    \begin{eqnarray}
\Phi_h: (\mathcal{M}, {\mathfrak{{g}}})\longrightarrow (\mathcal{M}, \mathfrak{g}'),\end{eqnarray}
what demonstrates that the 'trick' of relabeling $x'$ by $x$ can be understood as the isomorphism $\Phi_h$.

\

{
Finally, at the tensor level, we can see that the hole-diffeomorphism $\phi_h$ satisfies
$\phi_h^*\mathfrak{T}_{\mathfrak{g}}= \mathfrak{T}_{\phi_h^*\mathfrak{g}}$. Consequently,  the Einstein equation $\mathfrak{G}_{\mathfrak{g}}=
\kappa \mathfrak{T}_{\mathfrak{g}}$ transforms via $\phi_h^*$ to
$\mathfrak{G}_{\phi_h^*\mathfrak{g}}=
\kappa \mathfrak{T}_{\phi_h^*\mathfrak{g}}$, indicating that if $\mathfrak{g}^0$ is solution of 
$\mathfrak{G}_{\mathfrak{g}}=
\kappa \mathfrak{T}_{\mathfrak{g}}$, then $\phi_h^*\mathfrak{g}^0$ is also solution of the same equation.
}

\subsection{Hilbert's view of the Hole Argument}

As previously discussed, Einstein's interpretation of the hole argument revolves around a boundary problem that can be reformulated as an initial value problem.

\

In concise terms, this presents a version of Hilbert's hole argument as a Cauchy problem: in $\{x\}$-coordinates, consider an initial value spacelike surface defined by $x^4=\text{const.}$ with specific data (refer to \cite{ADM} for insights into the Hamiltonian approximation to General Relativity and \cite{Eric,Isenberg} for its application to the Cauchy problem in General Relativity). Assume that the evolution of the stress-energy tensor is predetermined, meaning its form is known throughout the Lorentz manifold. Subsequently, we seek to determine the metric as 
an evolution problem in time 
 $x^4$. Through a coordinate transformation that relocates points in the future of the initial value surface $x^4=\text{const.}$—while leaving the initial surface and the complementary of an open subset in the future, where the stress-energy tensor vanishes, unaffected—we find that the Einstein equations fail to uniquely determine the future. This is because the newly transformed metric in these coordinates is equally valid as a solution to the same field equations in the original coordinate system. Consequently, we can infer that the initial value problem lacks a unique solution in General Relativity.

\subsection{Physical interpretation of the Hole Argument}

To overcome the Hole Argument, Einstein approached it in a relational manner, asserting that the events of a manifold must be defined as coincidence points. Specifically, let $\widetilde{\gamma}_1$ and $\widetilde{\gamma}_2$ be two geodesics in $(\mathcal{M},\widetilde{\mathfrak{g}})$ intersecting at the point $p$ (a coincidence). Both geodesics are transformed by a diffeomorphism $\phi$ into two other ones ${\gamma}_1$ and ${\gamma}_2$ in $(\mathcal{M},{\mathfrak{g}}={\phi^{-1}}^*\widetilde{\mathfrak{g}})$, intersecting at the point $\phi(p)$.

The key question is not the value of a measure at a point $p$ of the pseudo-Riemann manifold $(\mathcal{M}, \widetilde{\mathfrak{g}})$ or at the same point of $(\mathcal{M}, {\mathfrak{g}})$. Instead, the meaningful question is about the value of a measure at the coincidence points. These coincidence points in both pseudo-Riemann manifolds are respectively $p$ and $\phi(p)$, and clearly, the same measure at both points leads to the same result. I refer the reader to Rovelli's book \cite{Rovelli}, specifically to Section 2.2.5 where Figure 2.4 vividly illustrates the meaning of coincidence points.

\

\

In conclusion, given the Einstein equation $\mathfrak{G}_{\mathfrak{g}}=\kappa \mathfrak{T}_{\mathfrak{g}}$, a hole-diffeomorphism transforms this equation into $\mathfrak{G}_{\phi_h^*\mathfrak{g}}=\kappa \mathfrak{T}_{\phi_h^\mathfrak{g}}$. Thus, following the point-coincidence interpretation, the solutions $\mathfrak{g}^0$ and $\phi_h^*\mathfrak{g}^0$ represent the same physical reality.

In the context of Gauge Covariance, i.e., assuming $\phi^*\mathfrak{T}_{\mathfrak{g}}= \mathfrak{T}_{\phi^*\mathfrak{g}}$, one can employ a rather complicated mathematical apparatus to present the relationalistic resolution of the Hole Argument.

In fact, one must contemplate solutions to the Einstein equation within the space $ \mathfrak{Q}_2(\mathcal{M})$, with the field equation taking the form:
\begin{eqnarray}\label{Einstein_equation1}    \mathfrak{G}_{[\mathfrak{g}]}=\kappa \mathfrak{T}_{[\mathfrak{g}]}.
\end{eqnarray}

Solving this equation is equivalent to solving, utilizing a Numerical Relativity scheme \cite{Eric}, the equation: \begin{eqnarray}
\mathfrak{G}_{\mathfrak{g}}=\kappa{\mathfrak{T}}_{\mathfrak{g}},
 \end{eqnarray}
 where, one might assume that $\mathcal{M}$ is globally hyperbolic, meaning it contains a Cauchy hyper-surface ${\Sigma}$.

More precisely, 
 choose  a normal field $N\in \mathfrak{X}(\mathcal{M})$ on $\Sigma$, that is, satisfying: 
\begin{eqnarray}
    \mathfrak{g}(N,N)=-1, \quad 
   \mathfrak{g}(N,X)=0 \quad \forall X\in \mathfrak{X}(\Sigma).\end{eqnarray}

Next, define the extrinsic curvature tensor in $\Sigma$,  as the $2$-covariant tensor 
$\mathfrak{K}:\mathfrak{X}(\Sigma)\times\mathfrak{X}(\Sigma)\rightarrow \mathcal{C}^{\infty}(\Sigma)$
satisfying:
\begin{eqnarray}
   \mathfrak{K}(X,Y)=-\mathfrak{g}(X,\nabla_Y N),\quad 
    \forall X,Y \in \mathfrak{X}(\Sigma),
\end{eqnarray}
and consider an orthonormal basis 
$\{E_j\}_{j=1,2,3}$ on $\mathfrak{X}(\Sigma)$, that is, 
$\mathfrak{g}(E_i,E_j)=\delta_{ij}$. 

Then,  Einstein's equation leads to four constraints in $\Sigma$:
\begin{eqnarray}\label{constraints}
    \mathfrak{G}_{\mathfrak{g}}(N,E_i)=\kappa {\mathfrak{T}}_{\mathfrak{g}}(N,E_i), \quad
    \mathfrak{G}_{\mathfrak{g}}(N,N)=\kappa {\mathfrak{T}}_{\mathfrak{g}}(N,N),\end{eqnarray}
which using the Gauss-Codazzi equations
\begin{eqnarray}
    R+2\mathfrak{Rc}(N,N)=R_{\mathfrak{h}}+(Tr_{\mathfrak{h}}\mathfrak{K})^2
    -\mathfrak{K}(E_i,E_j)\mathfrak{K}({E^i}^{\sharp}, {E^j}^{\sharp}),
    \nonumber\\
    \mathfrak{Rc}(N,E_i)=
(D^{\mathfrak{h}}_{E_i}\mathfrak{K})(E_j,{E^j}^{\sharp}) 
-(D^{\mathfrak{h}}_{E_j}\mathfrak{K})(E_i, {E^j}^{\sharp}),
\end{eqnarray}
where 
$\mathfrak{h}= \mathfrak{g}_{|_{\Sigma}}$ is the restriction of the metric $\mathfrak{g}$ to $\Sigma$,
$R_{\mathfrak{h}}$ is the Ricci scalar on $\Sigma$
and $D^{\mathfrak{h}}$ is the Levi-Civita connection in $\Sigma$, the constrains become:
\begin{eqnarray}\label{constraints1}
    R_{\mathfrak{h}}+(Tr_{\mathfrak{h}}\mathfrak{K})^2
    -\mathfrak{K}(E_i,E_j)\mathfrak{K}({E^i}^{\sharp},{E^j}^{\sharp})=
    2\kappa {\mathfrak{T}}_{\mathfrak{g}}(N,N)
    \quad \mbox{and} \quad
(D^{\mathfrak{h}}_{E_i}\mathfrak{K})(E_j,{E^j}^{\sharp}) 
-(D^{\mathfrak{h}}_{E_j}\mathfrak{K})(E_i, {E^j}^{\sharp})=\kappa {\mathfrak{T}}_{\mathfrak{g}}(E_i,N).\end{eqnarray}

\

What is crucial to realize here is that the extrinsic curvature tensor and, when fixing the values of ${\mathfrak{T}}_{\mathfrak{g}}(N,N)$ and ${\mathfrak{T}}_{\mathfrak{g}}(E_i,N)$ on $\Sigma$, the constraint equations (\ref{constraints1}) depend solely on the metric $\mathfrak{h}$.

\

With these considerations, the Einstein equation becomes an evolution problem with initial data $(\Sigma, \mathfrak{h}, \mathfrak{K})$ and fixed values of the stress-energy tensor on $\Sigma$, satisfying the constraints (\ref{constraints1}). If the initial value problem is well-posed (Ivonne Choquet-Bruhat proved this for the vacuum case \cite{Choquet}, but as far as I know, such a result does not exist in the general case), it would lead, up to diffeomorphisms, to a unique metric ${\mathfrak{g}}^0$. Assuming Gauge Covariance, the unique solution of (\ref{Einstein_equation1}) is $[{\mathfrak{g}}^0]$.

\subsubsection{Meaning of the coordinates}

In pre-relativistic mechanics, the $4$-dimensional manifold takes the form $\mathcal{M}=\mathcal{M}_3\times \mathbb{R}$, where the space-manifold $\mathcal{M}_3$ is equipped with a fixed metric $\mathfrak{g}_3$ that imparts meaning to any coordinate systems employed. This approach holds true in the special theory of relativity as well, where spacetime is represented by the Minkowski manifold. In both cases, the metric structure  of the manifold plays a crucial role in defining the physical framework and interpreting the coordinates within it. 
Therefore,  
the adoption of a fixed metric provides a consistent and meaningful foundation for understanding the geometry and physics associated with the respective pre-relativistic theories.

\

However, as we have already discussed, in General Relativity,
a metric determined as the solution of the Einstein equation can be transformed by an active diffeomorphism into an equivalent one, placing them within the same orbit.

\

Then, to elucidate the significance of a coordinate system within this  metric framework, we turn to a poignant statement by Misner, as quoted in \cite{Macdonald}:

{\it The metric [after a Gauge choice] defines not only the gravitational field that is assumed, but also the coordinate system in which it is presented. There is no other source of information about the coordinates apart from the expression for the metric. It is also not possible to define the coordinate system in any way that does not require a unique expression for the metric. In most cases where the coordinates are chosen for computational convenience, the expression for the metric is the most efficient way to communicate clearly the choice of coordinates that is being made.}

\

This is the key to comprehending Einstein's original Hole Argument. As elucidated earlier, we consider two distinct charts, denoted as $(\mathcal{M}, \psi_h)$ and $(\mathcal{M}, \psi'h)$, where $\psi_h$ and $\psi'_h$ coincide outside the hole but exhibit differences within it. Employing the technique of relabeling the $\{x'\}$ coordinates to $\{x\}$, Einstein discovers two distinct solutions, $g_{ij}(x)$ and $g'_{ij}(x)$, to the equations in $\{x\}$-coordinates.
However, as we have observed, this relabeling is essentially a means of effecting an active hole-diffeomorphism, denoted as $\phi_h=\psi_h^{-1}\circ \psi'_h$, allowing the expression of the metrics $\mathfrak{g}'$ and $\mathfrak{g}=\phi_h^*\mathfrak{g}'$ in terms of $\{x\}$-coordinates. Thus, in accordance with Misner's perspective, the physical interpretation of the $\{x\}$-coordinates differs when describing $(\mathcal{M}, \mathfrak{g})$ and $(\mathcal{M}, \mathfrak{g}')$.

\

Following Davis's line of thinking depicted in \cite{Davis1}, in General Relativity, the metric is only available to define a Lorentz manifold after the Einstein equations are solved. However, the field equations have to be solved using coordinates of unknown physical meaning. This endorses a privileged role to each metric solution, consisting of determining the physical meaning of the ${x}$-coordinates in terms of which it is written.

Well understood, different metric solutions to the field equations may give different physical meanings to the same set of local coordinates. The fact that the ${x}$-coordinates used to find a metric $g_{ij}(x)$ are the same quadruple of real numbers to find another solution ${g}'_{ij}(x)$ does not mean that the ${x}$-coordinates have the same physical meaning in both solutions. Each metric solution brings its own assignment of physical or geometrical meaning to the local coordinates used to write it.

\

In a few words, as stated by Davis: {\it Coordinates have no independent meaning, independent of the metric solution. Each of the multiple metric solutions carries its own physical interpretation of its own local coordinates}. The example of the Schwarzschild metric studied in \cite{Macdonald} is very illuminating and helps to clearly understand what was mentioned earlier.

\section{Some remarks on the philosophical viewpoint of the Hole Argument}

First of all, and this is my opinion, there are trees (mathematical structures) that, unfortunately, does not always lead some authors to see all  the forest (its physical implications). Even worse, some researchers do not look at the trees with the correct glasses, often getting entangled in the intricate branches and details without grasping the broader implications for the physical reality they seek to understand. Let me elaborate on this observation:

\

In delving into the intricacies of mathematical formalism, it is crucial to maintain a clear perspective on the overarching physical principles and concepts that these mathematical structures are meant to capture. It is akin to navigating through a dense forest of mathematical symbols and equations—while the details of individual trees (a concrete mathematical apparatus) are undoubtedly important, it is equally essential to step back and appreciate the collective landscape they form (physical implications). Failure to do so may result in a disconnect between the elegance of mathematical formulations and their true significance in describing the underlying physics.

\

In this context, my aim is to analyze certain works associated with the Hole Argument. For instance, in \cite{Earman-Norton}, the authors characterize a model, denoted as $T$ in a given theory, as an $n+1$ tuple $(\mathcal{M}, O_1, \ldots, O_n)$, where $O_j$ represents geometric objects defined on $\mathcal{M}$ that adhere to specific tensor equations, namely $O_k=0, \ldots, O_n=0$, where $k$ is less than $n$.

The authors articulate the Gauge Theorem in the following manner: "If $(\mathcal{M}, O_1, \ldots, O_n)$ represents a model of a theory and $\phi$ is a diffeomorphism of $\mathcal{M}$ onto itself, then $(\mathcal{M}, \phi^*O_1, \ldots, \phi^*O_n)$ also constitutes a model of the theory." Subsequently, the authors proceed to establish this theorem in an unconventional manner, employing coordinates.

Essentially, their objective is to demonstrate that the vanishing equations $O_k=0, \ldots, O_n=0$ are "preserved by diffeomorphisms." However, this proof appears trivial, as (assuming, for simplicity, that the equations are given by $2$-covariant tensors), for any $j=k, \ldots, n$, one has
\begin{eqnarray}
    \phi^*O_j(X,Y)=O_j(\phi_*X,\phi_*Y)=0,
\end{eqnarray}
because $O_j$ vanishes for all pairs of vector fields.

\

Here, the crucial point is that this does not imply that the "carry along" of the tuple $(\mathcal{M}, \phi^*O_1, \ldots, \phi^*O_n)$ is inherently a model of the theory. I have previously addressed this matter in Section \ref{gauge-invariance} following the introduction of the concept of Gauge Covariance.
A clear understanding of Gauge Covariance implies, particularly in terms of a simple stress-energy tensor $\mathfrak{T}_{\mathfrak{g}}$ (with the general definition provided in \ref{gauge-invariance}),
\begin{eqnarray}
\phi^*\mathfrak{T}_{\mathfrak{g}}=\mathfrak{T}_{\phi^*\mathfrak{g}}\Longleftrightarrow
(\phi^*\mathfrak{T}_{\mathfrak{g}})(X,Y)= \mbox{(by definition)}
=\mathfrak{T}_{\mathfrak{g}}(\phi^*X,\phi^*Y)=\mathfrak{T}_{\phi^*\mathfrak{g}}(X,Y)\quad \forall X,Y\in \mathfrak{X}(\mathcal{M}).\end{eqnarray}

\

In my view, the scenario in \cite{Earman-Norton} is characterized by non-vanishing fields $\mathfrak{A}_1, \ldots, \mathfrak{A}_k$, while the vanishing ones $O_j=0$ represent equations such as $\mathfrak{B}_j(\mathfrak{g}, \mathfrak{A}_1, \ldots, \mathfrak{A}_k)=0$. Consequently, Gauge Covariance simply signifies:
\begin{eqnarray}\label{abcd}
\phi^*\mathfrak{B}_j(\mathfrak{g},\mathfrak{A}_1,...,\mathfrak{A}_k)=
    \mathfrak{B}_j(
\phi^*\mathfrak{g},\phi^*\mathfrak{A}_1,...,\phi^*\mathfrak{A}_k),   \end{eqnarray}
and therefore, $(\mathcal{M}, \phi^*O_1,..., \phi^*O_n)$ also qualifies as a model of the theory if the equations governing the theory remain invariant under diffeomorphisms.

{

This fact is crucial because the statement provided in \cite{Earman-Norton}, which asserts that given a model $(\mathcal{M}, O_1, \ldots, O_n)$ and $\phi$ a diffeomorphism, then $(\mathcal{M}, \phi^*O_1,..., \phi^*O_n)$ is also a model, is fundamentally incorrect. This assertion holds only if the equations $O_j=0$ are Gauge Covariant in the sense of (\ref{abcd}). If it is not, then the equations represent a different physical situation, and thus, one cannot argue that different solutions represent the same reality.

Einstein clearly understood this when formulating the Hole Argument, as he did not consider Gauge Covariance. For this reason, he introduced hole-diffeomorphisms, which are Gauge Covariant with respect to the stress-energy tensor. However, this crucial fact is not addressed at all in \cite{Earman-Norton}.

Consequently, if one assumes Gauge Covariance, the Hole Argument loses its meaning, as one can see}
in \cite{Earman-Norton}, where the Hole Argument is presented as a corollary, stated as follows: "Let $T$ be a model with a manifold $\mathcal{M}$ and $\mathcal{H}$ (the hole) be any neighborhood of $\mathcal{M}$. Then there exist arbitrarily many distinct models of the theory on $\mathcal{M}$ which differ from one another only within $\mathcal{H}$."

To substantiate this claim, the authors employ hole-diffeomorphisms—those that deviate from the identity only within the hole. They argue that there are arbitrarily many such pulled-back models that fulfill the specified requirement.
This highlights, as I have mentioned in Remark \ref{Remark}, that in a Gauge Covariant theory, the Hole serves as a mere tool to illustrate that choosing diffeomorphisms different from the identity yields another distinct model of the theory $T$. {
That is, the key point is that under Gauge Covariance, all the diffeomorphisms pull back models to models, and one does not need any kind of special diffeomorphisms. This can be clearly seen in the proof of the Hole corollary presented on page 523 of \cite{Earman-Norton}, where, in order to assume that the hole-diffeomorphisms pull back models to models, the authors use their "Gauge Theorem", which, if well understood, not as the authors do, is only the assumption of Gauge Covariance.}

\

Exactly, the crux of Einstein's Hole Argument lies in this fact—the invariance under certain diffeomorphisms (specifically, hole-diffeomorphisms in the absence of a Gauge Invariant theory). As I have previously detailed, with a hole diffeomorphism, one observes that $\phi_h^*\mathfrak{T}_{\mathfrak{g}}=\mathfrak{T}_{\phi_h^*\mathfrak{g}}$, implying that the Einstein equation remains unchanged. Consequently, for a given solution $\mathfrak{g}^0$, both $\mathfrak{g}^0$ and $\phi_h^*\mathfrak{g}^0$ stand as solutions to the same Einstein equation.

\

\

Furthermore, the authors of \cite{Earman-Norton} state: "The name of this corollary stems from Einstein's original discovery of it in a specialized form. He considered a matter-free hole in a source mass distribution and showed that the gauge freedom of any generally covariant gravitational field equation in general relativity allowed multiple metric fields within the hole." This, of course, encapsulates the true essence of the Hole Argument. However, in their formulation, the authors do not delve into the core of the argument, which lies in the vanishing of the stress-energy tensor—distinct from an intrinsic curvature tensor—in the hole. They directly assume Gauge Covariance, rendering the Hole inconsequential in their reasoning.

To wrap up, once one embraces the relationalist viewpoint through Einstein's coincidence points, all hole-diffeomorphisms define the same model. Essentially, the hole serves to bolster a substantivalist perspective against General Covariance. Expanding on this, if one assumes Gauge Covariance, then all diffeomorphisms define the same model. From a substantivalist standpoint, one can argue that the theory becomes indeterministic.

\

Having clarified the Hole Argument, it appears to me that some authors {continues employing} an inaccurate version of it. For instance, in \cite{Gomes}, the authors define General Covariance, in terms of models, as follows:

If $(\mathcal{M},\mathfrak{g},\mathfrak{T}_{\mathfrak{g}})$ is a model, and $\phi:\mathcal{M}\rightarrow\mathcal{M}$ is a diffeomorphism. Then, $(\mathcal{M},\phi^*\mathfrak{g},\phi^*\mathfrak{T}_{\mathfrak{g}})$  is also considered a model.

\

Clearly, this statement holds true only if one accepts Gauge Covariance, i.e., $\phi^*\mathfrak{T}_{\mathfrak{g}}=
\mathfrak{T}_{\phi^*\mathfrak{g}}$. Subsequently, the authors embark on a "mathematico-philosophical" discussion, employing mathematical concepts such as category theory to distinguish different models related via diffeomorphisms and arguing that they represent different possibilities. The issue here is that, in all their arguments, everything is intertwined with mathematics—algebraic structures, Lie derivatives, Noether's second theorem, considering $4$ space-times as leaves of a $5$ dimensional manifold, and more.

However, the crux of the matter is that there is a lack of physics in these discussions. While there is a reference to "Quantum reference frames," it is evidently unrelated to the Hole Argument. What I would emphasize is that to genuinely address the Hole Argument, one must incorporate physics, particularly Einstein's point-coincidences, which appears to be absent in \cite{Gomes}.

\

Another example is \cite{Curiel}, where the author presents the Hole Argument as an evolution problem, as follows:
{\it
"Fix an space-time model $(\mathcal{M}, g_{\mu\nu})$. For ease of exposition, we
stipulate that the space-time be globally hyperbolic, and so possesses a global Cauchy surface $\Sigma$.
Say that we know
the metric tensor on $\Sigma$ and on the entire region of space-time to its causal past.
It is known that this forms a well set Cauchy problem, and so there is a solution
to the Einstein field equation that uniquely extend $g_{\mu\nu}$ on the causal pass of $\Sigma$ to metric tensor  on all of $\mathcal{M}$,
yielding the original space-time we fixed. In particular, the solution to the Cauchy problem fixes the
metric on the region to the causal future of $\Sigma$."}

\

Next, the author considers a diffeomorphism $\phi$ that leaves the causal past of $\Sigma$ invariant. Subsequently, the author performs an unusual "mathematical" operation—applying $\phi^*$ to the metric but not to the manifold itself. This can be seen as a less developed version of the Einstein trick, where one typically changes $g'_{\mu\nu}(x')$ to $g'_{\mu\nu}(x)$. Then, the author assesses:

{\it This yields a seemingly different metric—a different “physical state of the gravitational field”- on the causal future of $\Sigma$, in the sense that the same points of the causal future of $\Sigma$ now carry (in general) a different value for the metric. This is the crux of the issue, that the diffeomorphism applied to the metric has yielded a different tensor field in the sense that the same points of the space-time manifold now carry a different metric tensor than before.}

\

Certainly, this appears to be a misinterpretation of the initial value problem. Firstly, the author specifies the values of the metric (and presumably those of the extrinsic curvature) on $\Sigma$ based on the metric across the entire manifold, without any mention of the stress-tensor. This suggests, to me, an assumption that they are dealing with the vacuum case. Utilizing the Choquet-Bruhat theorem \cite{Choquet}, one can establish that, for the vacuum Einstein equation, there exists only one extension of the metric, up to diffeomorphisms (the mathematical ones), to the entire manifold $\mathcal{M}$.

This contradicts the argument presented in \cite{Curiel}, as their "operation"—applying $\phi^*$ solely to the metric—yields another metric that is distinct and not related via diffeomorphisms to the original one. This contradicts the uniqueness guaranteed by the Choquet-Bruhat theorem \cite{Choquet} (see also \cite{Landsman, Isenberg}). In other words, the newly constructed metric in this manner is not a solution to Einstein's vacuum equation with the fixed initial conditions on $\Sigma$ provided by the original metric.

Once again, it seems that the author is dealing with two different problems, and consequently, the distinct solutions correspond to two different physical situations, avoiding indeterminism. (See also \cite{Arledge} for a critical review of \cite{Curiel}).

\

Another paper that approaches the Hole Argument in an unconventional manner is \cite{Weatherall} (see \cite{Pooley-Read} for a critical review of it). In this paper, the author explicitly utilizes the Earman-Norton {trivial} definition of the Hole Argument, implicitly assuming, {although not realizing it}, Gauge Covariance. The author then posits that isometric pseudo-Riemann manifolds are empirically equivalent, meaning that active diffeomorphisms (in the mathematical sense, not in the sense assumed in \cite{Weatherall}) provide the same physical reality. However, this argument is presented without employing a relationalism viewpoint to support it, indicating to me that the author may lack some depth in understanding the subject, particularly in relation to Einstein's resolution of the problem.

\

However, the author goes beyond and employs the same unusual "mathematical" reasoning, as explained earlier: applying $\phi^*$ to the metric but not to the manifold. The author compares $(\mathcal{M}, \mathfrak{g})$ with $(\mathcal{M}, \phi^*\mathfrak{g})$ at the same point $p$, which the author considers an "active" diffeomorphism in contrast to the mathematical definition (the author also uses the term "push-forward" to describe the pull-back operation and employs various incorrect mathematical concepts).

Once again, the use of the Hole seems to be a mere instrument to express that the diffeomorphism $\phi$ is not the identity, devoid of any real connection to the actual problem concerning General Covariance.

Regardless, the author concludes that in this case, $(\mathcal{M}, \mathfrak{g})$ and $(\mathcal{M}, \phi^*\mathfrak{g})$ are not equivalent. This conclusion, however, seems trivial, as using this approach, the Lorentz manifolds $(\mathcal{M}, \mathfrak{g})$ and $(\mathcal{M}, \phi^*\mathfrak{g})$ stem from different initial values on a Cauchy surface, as I have already explained in the context of \cite{Curiel}. Therefore, this viewpoint appears to be disconnected from the real Hole Argument, which necessitates a genuine active diffeomorphism to determine whether one adopts a substantival viewpoint and accepts indeterminism or rejects indeterminism by choosing the relationalist option, as Einstein did.

\

Finally, in my opinion, more metaphysical arguments like the ones presented in \cite{Fletcher}, discussed critically in \cite{Arledge}, seem to obscure the real debate about the Hole Argument. They do not contribute to visualizing its importance in the context of the ongoing debate about substantivalism vs. relationalism in the physical interpretation of the world.

\section{Conclusions}
In the course of { this essay}, I have observed the significance of the Hole—defined as a set where the stress-energy tensor vanishes—particularly in the context of General Covariance, which entails invariance with respect to changes in coordinates without assuming Gauge Covariance. It is noteworthy that, under the assumption of Gauge Covariance, the Hole ceases to play a role, as the equations maintain invariance under diffeomorphisms.

\

Additionally, I have delved into Einstein's original Hole Argument, revealing its connection to coordinates as a manifestation of Hole Argument on the manifold through hole-diffeomorphisms. From a physical standpoint, there is a compelling argument for resolving the Hole Argument "à la Einstein," which involves abandoning substantivalism and embracing a relationalistic perspective where events are defined by coincidences. Einstein's rejection of the substantivalism vision of space-time is succinctly expressed in his words: {\it The requirement of General Covariance takes away from space and time the last remnant of physical objectivity.}
Rovelli echoes this sentiment, stating: {\it Not anymore fields on space-time, just fields on fields.}

\

{
Throughout the essay, I have emphasized the unconventional treatment of the Hole Argument by some philosophers. It appears to me that they fail to recognize the importance of the stress-energy tensor, as opposed to the intrinsic curvatures of the manifold, which may not be preserved by certain pull-backs. This is precisely why Einstein introduces passive hole diffeomorphisms—to ensure that exactly the same equations are preserved. Consequently, this results in different metrics satisfying the same field equations. Initially, this led Einstein to reject General Covariance. However, he later realized, by removing substantivalism, that these different metrics define the same physical reality, allowing for the preservation of General Covariance in the theory.

Hence, when this physical perspective is overlooked, the Hole Argument becomes a mere artifact, used to implicitly support hidden Gauge Covariance, even if it is not always consciously recognized. For instance, in \cite{Earman-Norton}, authors use this argument to trivially demonstrate that the Gauge Covariant field equations yield different solutions. Subsequently, they delve into a philosophical-mathematical debate that, from a physical standpoint, seems irrelevant.

\

I have endeavored to elaborate on this concept in Section V, expressing my reservations about an overreliance on intricate geometric structures such as fiber bundles, morphisms, categories, homotopy theory, spin networks, $C^*$-algebras, rings, or functors within the realm of physics (see, for instance, \cite{Ladyman, Halvorson, Rickle, Yaghmaie, Bain} and the references therein). From my perspective as a cosmologist and mathematician, such an approach risks losing sight of the fundamental physical essence of the problem. To be more precise, I advocate for a return to the core of the issue: "Less math, just physics."}

\section*{Acknowledgments} 
This work  
is supported by the Spanish grant 
PID2021-123903NB-I00
funded by MCIN/AEI/10.13039/501100011033 and by ``ERDF A way of making Europe''.

\end{document}